\documentclass[12pt,aps,prd,reprint,roupedaddress,showpacs,nofootinbib]{revtex4-1}
\usepackage{amsmath,amssymb,dsfont,mathrsfs,graphicx,float,caption,subcaption,epstopdf}
\usepackage{ragged2e}
\usepackage[bottom]{footmisc}
\DeclareMathOperator{\Tr}{Tr}

\begin{document}


\title{Testing modified gravity and no-hair relations for the Kerr-Newman metric through quasi-periodic oscillations of galactic microquasars}


\author{Arthur George Suvorov}
\email{suvorova@student.unimelb.edu.au}
\affiliation{School of Physics, University of Melbourne, Parkville VIC 3010, Australia}
\author{Andrew Melatos}
\email{amelatos@unimelb.edu.au}
\affiliation{School of Physics, University of Melbourne, Parkville VIC 3010, Australia}


\date{\today}

\begin{abstract}

We construct multipole moments for stationary, asymptotically flat, spacetime solutions to higher-order curvature theories of gravity. The moments are defined using $3+1$ techniques involving timelike Killing vector constructions as in the classic papers by Geroch and Hansen.
Using the fact that the Kerr-Newman metric is a vacuum solution to a particular class of $f(R)$ theories of gravity, we compute all its moments, and find that they admit recurrence relations similar to those for the Kerr solution in general relativity. 
It has been proposed previously that modelling the measured frequencies of quasi-periodic oscillations from galactic microquasars enables experimental tests of the no-hair theorem.
We explore the possibility that, even if the no-hair relation is found to break down in the context of general relativity, there may be an $f(R)$ counterpart that is preserved. We apply the results to the microquasars GRS $1915$+$105$ and GRO J$1655$-$40$ using the diskoseismology and kinematic resonance models, and constrain the spins and `charges' [which are not really electric charges in the $f(R)$ context] of their black holes.

\end{abstract}

\pacs{04.50.Kd, 04.40.Nr, 98.54.Aj, 03.50.-z }	

\maketitle

\section{Introduction}
The interpretation of parameters appearing within solutions to the theory of general relativity (GR) has a rich history \cite{gripod,bicak}. In linear field theories, such as Newtonian gravity or Maxwell electrodynamics, it is well understood how the potential associated with a solution to the field equations can be characterised by its multipole moments \cite{jannew,gerochflat}. The moments themselves generate a series expansion for the field potential, which then offers a term-by-term physical understanding \cite{raab}. Nonlinear field theories, such as GR, do not obey the principle of superposition, making multipole decomposition harder to define. Some of the better known multipole expansions in GR are due to Thorne \cite{thorne}, Curtis \cite{curtis}, Geroch \cite{gerochcurvedmult}, Hansen \cite{hansen}, Janis and Newman \cite{jannew} and others \cite{beigsim,hoensel,simon}.

One practical application of multipole moments is to characterise solutions in non-GR theories \cite{glampbab,papsot}, e.g. $f(R)$ gravity \cite{felice}. Interestingly there exist vacuum solutions to higher-order curvature theories which are simultaneously nonvacuum solutions in GR. For example, the Kerr-Newman metric \cite{kerrnew} is a \emph{vacuum} solution to a particular class of $f(R)$ gravities \cite{cruzd,larr,cembranos}. In the GR context, astrophysical interest in the Kerr-Newman solution is limited due to the expectation that naturally formed black holes are neutral. However, in $f(R)$ gravity, the Kerr-Newman metric serves as a natural extension of the uncharged Kerr solution. In this paper, we explore the properties of this metric as a solution to a vacuum $f(R)$ theory. We introduce multipole moments for higher-order curvature theories through the Geroch-Hansen procedure, \cite{gerochcurvedmult,hansen,gerochgen,gerochflat,gerochgen2} compute the Kerr-Newman moments, and find that they reduce to their Kerr counterparts found by Hansen, \cite{hansen} when the appropriate limit is taken.

How can one determine if the Kerr-Newman solution in $f(R)$ gravity occurs in Nature? A contender for an experiment to answer this question is provided by quasi-periodic oscillations (QPOs) in X-ray binaries \cite{lambetal,watts} or active galactic nuclei \cite{gierl}. While the physical origin of these phenomena is still being debated \cite{watts}, some models suggest that, as the accretion disk swirls around the black hole, certain fluid modes become trapped within the ergosphere \cite{tchek}, producing quasi-periodic patterns of electromagnetic activity via the Lense-Thirring effect \cite{wagoner,ingram}. In GR, owing to the uniqueness results of Israel, Carter, Robinson, and Hawking \cite{chandbh,heusler,hawkel}, the frequencies of these modes are directly related to the mass and spin of the Kerr black hole. Johannsen and Psaltis \cite{johpsa} have demonstrated methods by which, given an independent measurement of either the mass or spin, one can determine whether the no-hair theorem in GR is violated by monitoring QPO frequencies, or whether something else is going on, e.g. $f(R)$ gravity.

The multipole relations we find for the Kerr-Newman solution in $f(R)$ gravity obey recurrence relations similar to those of the Kerr metric in GR, in that each moment is computable from the lowest-order mass and current moments. This result indicates the presence of a no-hair-type result in $f(R)$ gravity, except with three parameters instead of two. Although it is unlikely that the Kerr-Newman metric is the unique descriptor of black holes in $f(R)$ gravity, it may form part of an extended family of solutions generated by the function $f$. This presents an alternative scenario to that presented by Johannsen and Psaltis \cite{johpsa}: if no-hair-violating data are discovered, the no-hair theorem may still be preserved, provided that GR does not apply. Hence, for any given object, an independent mass measurement combined with a QPO pair may not be enough to rule out no-hair relations altogether. More generally, one would need $m$ measurements to potentially invalidate a no-hair theorem involving $m-1$ independent parameters, but one would need only three measurements to determine whether GR (or the Kerr no-hair relation) is consistent.

Given a QPO pair, one can compare theoretical relations between the mass, spin, and quadrupole moment with measured frequencies. The Kerr-Newman metric in $f(R)$ gravity contains a parameter $\rho$ (no longer charge). Given the measured frequencies of QPO pairs from the microquasars GRS $1915$+$105$ and GRO J$1655$-$40$, we place bounds on $\rho$ following Johannsen and Psaltis \cite{johpsa}. Some GR estimates place the spin of GRS $1915+105$ as high as $a/M = 0.998$ \cite{middle}, where $f(R)$ effects may be strong \cite{capbeyond,glampbab,remillard}. A similar analysis using the Kerr-Newman metric indicates that the spin of these objects may be over or underestimated depending on the sign of $\rho$.


The paper is structured as follows. In $\S 2$ we revisit the definitions given by Geroch and Hansen in GR and generalise them to alternative gravities. In $\S 3$ we derive the equations which govern the multipole moments for spacetimes in general gravitational theories. In $\S 4$ we explicitly solve these equations for the $f(R)$ Reissner-N{\"o}rdstrom and Kerr-Newman black holes and use the solutions to generate the multipole moments. In $\S 5$ we explore the consequences for QPO modelling to illustrate how the moment calculation can be applied astrophysically. We discuss the results in $\S 6$.

\section{Geroch-Hansen moments In GR}

\subsection{Definitions}
The Geroch and Hansen moments \cite{gerochcurvedmult,hansen} are defined for asymptotically flat\footnote{We assume a 4-dimensional spacetime throughout and zero cosmological constant in GR.}, stationary\footnote{A spacetime is \emph{stationary} if the metric tensor does not depend on any time-like coordinates.} solutions to the Einstein equations outside of a source,
\begin{equation} \label{eq:ein}
R_{\mu \nu} = 0  ,
\end{equation}
where $R_{\mu \nu}$ denotes the Ricci tensor. They are built out of potentials associated with the metric which satisfy Poisson-like equations, thereby relinearizing a significant portion of the problem. The geometric framework that facilitates the relinearization, described below, also provides a convenient recipe for generalising from the Einstein theory to alternative gravitational theories.

We assume that the spacetime $M$ is stationary, so that there exists a timelike vector field $\boldsymbol\xi$ \cite{wald} which satisfies Killing's equation\footnote{We consider torsionless theories of gravitation:  $\nabla_{\mu} g_{\alpha \beta} = 0$.}
\begin{equation} \label{eq:killing}
\nabla_{\mu} \xi_{\nu} + \nabla_{\nu} \xi_{\mu} = 0 .
\end{equation}
If we define $\lambda$, the norm of $\boldsymbol\xi$, by \cite{gerochgen}
\begin{equation} \label{eq:norm}
\lambda = \xi^{\alpha} \xi_{\alpha} ,
\end{equation}
and the twist, $\boldsymbol{\omega^{\dag}}$, by \cite{wald}
\begin{equation} \label{eq:twist}
\omega^{\dag}_{\alpha} = \epsilon_{\alpha \beta \gamma \delta} \xi^{\beta} \nabla^{\gamma} \xi^{\delta} ,
\end{equation}
then the metric $\boldsymbol{g}$ may be written in the generalised Papapetrou form\footnote{Throughout, Greek symbols range over spacetime indices $0,1,2,3$ while Latin indices are reserved for spatial indices $1,2,3$.} \cite{beigsim,chandbh}
\begin{equation} \label{eq:papapet}
ds^2 = \lambda ( dt + \sigma_{i} dx^{i} )^{2} - \lambda^{-1} h_{ij} dx^{i} dx^{j} ,
\end{equation}
where $\omega^{\dag}_{i} = - \lambda^{-2} \epsilon_{ijk} D^j \sigma^{k}$, and $D$ forms the covariant derivative with respect to $h_{ij}$ (see Appendix A). The spacetime then admits a $1+3$ split, and we denote the manifold associated with the Riemannian metric $\boldsymbol{h}$ as $S$ \cite{hansen,beigsim,simonbeig}. We assume that $S$ is simply-connected (such as the exterior region to a black hole ergosphere), to ensure the existence of a certain scalar field (see below) \cite{beigsim,brezis}.

\subsection{Static and stationary spacetimes}

Geroch initially considered static spacetimes with $\boldsymbol{\omega^{\dag}} = 0$ [2,3]. Starting from Einstein's equations on $M$, $R_{\mu \nu} = 0$, he derived the corresponding field equations over $S$. The fundamental variables to be computed are $\lambda$ and the components $h_{ij}$. The equation for $\lambda$ is equivalent to a conformally invariant Laplace equation for a particular scalar field $\phi^{M}$ (mass potential), for which the moments are defined. The benefit of trading $\lambda$ for $\phi^{M}$ is that, when the spacetime is Minkowski $(\boldsymbol{g} = \boldsymbol{\eta})$, the Newtonian equation $\nabla^2 \phi^{M} = 0$ is recovered. In particular, the procedure uniquely pulls out the mass monopole moment for the Schwarzschild black hole, with all other moments being identically zero.

The extension to the stationary case $(\boldsymbol{\omega^{\dag}} \neq 0)$ was given by Hansen \cite{hansen}. For example, rotating solutions require an additional moment-generating field, $\phi^{J}$ \cite{heusler,ryan}. Combining \eqref{eq:killing} and \eqref{eq:twist} with the Bianchi identities leads to
\begin{equation} \label{eq:curl}
\nabla_{\alpha} \omega^{\dag}_{\beta} - \nabla_{\beta} \omega^{\dag}_{\alpha} = -\epsilon_{\alpha \beta \gamma \delta} \xi^{\gamma} R^{\delta}_{\tau} \xi^{\tau}  .
\end{equation}
Assuming the Einstein condition \eqref{eq:ein}, the right-hand side of \eqref{eq:curl} vanishes, and there exists a scalar field $\omega^{\dag}$ such that 
\begin{equation} \label{eq:omegadagger}
\nabla_{\alpha} \omega^{\dag} = \omega^{\dag}_{\alpha} .
\end{equation}
We explore generalisations of the quantity $\boldsymbol{\omega^{\dag}}$ in the next section to extend the formalism to non-Einstein theories, since the scalar $\omega^{\dag}$ does not exist in general outside GR.

To ensure that the spacetime $M$ is asymptotically flat, one requires the existence of a manifold $\tilde{S}$, consisting of $S$ plus one additional point $\Lambda$, subject to certain conditions \cite{gerochcurvedmult,gerochgen}. This point $\Lambda$ is to be thought of as the `asymptotic' point, in the sense that we can extend the metric $\boldsymbol{h}$ to the boundary of $S$. In particular, we require the existence of a scalar field, $\Omega$ such that $\tilde{h}_{a b} = \Omega^2 h_{a b}$ is a metric on $\tilde{S}$, and at $\Lambda$ one has $\Omega = 0$, $\tilde{D}_{a} \Omega = 0$, and $\tilde{D}_{a} \tilde{D}_{b} \Omega = n \tilde{h}_{ab}$ for some $n$. Euclidean $3$-space is asymptotically flat in the above sense with conformal factor $\Omega = r^{-2}$, where $r$ is the distance from some origin \cite{gerochcurvedmult,hansen,gerochgen}.

One can now express the field equations on $M$ in terms of the variables $\{\lambda,\omega^{\dag},h_{ij}\}$. In Appendix A we show how to map these equations onto $S$. The equations of motion for $\lambda$ and $\omega^{\dag}$ are replaced by ones for $\phi^{M}(\lambda,\omega^{\dag})$ and $\phi^{J}(\lambda,\omega^{\dag})$ on $S$ and $\tilde{S}$, which read\footnote{Scripted letters refer everywhere to curvature tensors on $S$.} \cite{gerochcurvedmult,hansen}
\begin{equation} \label{eq:poisson}
(D^{i} D_{i} -\tfrac{1}{8} \mathscr{R}) \phi^{A} = \tfrac{15}{16} \lambda^{-2} \kappa \phi^{A} ,
\end{equation}
with $A = M,J$, where $\mathscr{R}$ is the Ricci scalar on $S$, and $\kappa = \left( D^{m} \lambda D_{m} \lambda + D^{m} \omega D_{m} \omega \right)$. Equation \eqref{eq:poisson} is solved subject to mixed Dirichlet and Neumann boundary conditions (see $\S 3$). The factor $\tfrac {1} {8} \mathscr{R}$ ensures that the field equations are conformally invariant with respect to $\phi^{A}$ (see Appendix B), so that transforming into the space $\tilde{S}$ leaves the moments unaltered \cite{harris,beigsim}. 

We remark in passing that one can work within the original formalism due to Geroch \cite{gerochcurvedmult}, where equation \eqref{eq:poisson} is taken to be homogenous $(\kappa = 0)$, without altering the results of this paper. The inhomogenous terms added by Hansen (equations (2.16) in \cite{hansen}) preserve the conformal properties of the Laplace equation while permitting closed form solutions $\phi^{A}$ to \eqref{eq:poisson} in GR. However, this calculational advantage does not carry over to modified gravities except in special cases. When the Ricci tensor is complicated, no closed-form solutions exist in general, as there is no obvious way to write the derivatives of $\omega$, $\boldsymbol{Q}$, or $\lambda$ in terms of the 3-metric $h_{ij}$ [see \eqref{eq:ric}]. The homogeneous and inhomogenous forms of the non-GR extension of equation \eqref{eq:poisson} both produce the same moments, as calculated in $\S 4$B, according to Theorem 4 in \cite{simonbeig}, even though the generalised $\phi^{A}$ are different in the two cases. In this paper, we persevere with Hansen's inhomogenous formalism to match equation (2.16) in \cite{hansen}, at the cost of mildly complicating the form of equation \eqref{eq:poisson}.

In order to define the moments, we construct recursively a set of tensor fields $P^{A}_{a_{1} \cdots a_{s}}$ on $(\tilde{S}, \boldsymbol{\tilde{h}})$ by
\begin{equation} \label{eq:multmom}
\begin{split}
&P^{A} = \tilde{\phi}^{A} , \\
&P^{A}_{a_{1} \cdots a_{s+1}} = \mathscr{C} \big[ \tilde{D}_{a_{1}} P^{A}_{a_{2} \cdots a_{s +1}} - \frac {1} {2} s (2 s -1) \mathscr{\tilde{R}}_{a_{1} a_{2}} P^{A}_{a_{3} \cdots a_{s+1}} \big] ,
\end{split}
\end{equation}
with $\tilde{\phi^{A}} = \Omega^{-1/2} \phi^{A}$, where $\mathscr{C} [J_{a \cdots b}]$ denotes the symmetric, trace-free part of $J_{a \cdots b}$ and $\mathscr{\tilde{R}}_{ij}$ is the Ricci tensor associated with $\tilde{h_{ij}}$. We now define the $2^{s}$ moment of $\phi^{A}$ to be the value of $P^{A}_{a_{1} \cdots a_{s}}$ at $\Lambda$ and write $P^{M}_{a_{1} \cdots a_{s}} = M_{a_{1} \cdots a_{s}}$ and $P^{J}_{a_{1} \cdots a_{s}} = J_{a_{1} \cdots a_{s}}$.

For some physical intuition, consider the Newtonian gravitational potential $\boldsymbol{\phi}$ defined on Euclidean 3-space. To perform the conformal completion we take $\Omega = r^{-2}$ and $\tilde{\boldsymbol{\phi}} = \Omega^{-1/2} \boldsymbol{\phi} = r \boldsymbol{\phi}$. The vacuum Newtonian gravitational potential [satisfying \eqref{eq:poisson}] admits an expansion of the form \cite{hansen}
\begin{equation} \label{eq:newtmultmom2}
\tilde{\boldsymbol{\phi}} = U + U_{a} x^{a} + \frac {1} {2!} U_{a b} x^{a} x^{b} + \cdots \,\,\, ,
\end{equation}
where $\boldsymbol{U}$ are the multipole moments and $\boldsymbol{x}$ is the position vector. It follows immediately that at the point $\Lambda$ $(|\boldsymbol{x}| = 0)$ we have
\begin{equation} \label{eq:newt}
U = \tilde{\boldsymbol{\phi}}|_{\Lambda}, \,\,\,  U_{a} = \tilde{D}_{a} \tilde{\boldsymbol{\phi}}|_{\Lambda}, \,\,\,  U_{ab} = \tilde{D}_{a} \tilde{D}_{b} \tilde{\boldsymbol{\phi}}|_{\Lambda}, \cdots \,\,\, .
\end{equation}
Equation \eqref{eq:newt} coincides exactly with equation \eqref{eq:multmom} when the curvature terms vanish. 

In GR, one may solve \eqref{eq:poisson} exactly for any stationary vacuum solution \cite{hansen}. For $R_{\mu \nu} \neq 0$, factors emerge involving the Ricci tensor, such as $R_{\mu \nu} \xi^{\mu}\xi^{\nu}$, which make finding a general solution difficult. One must solve the non-GR extended version of equation \eqref{eq:poisson}, namely equation \eqref{eq:nongrpois}, for individual metrics in any given modified theory of gravity to define the moments using \eqref{eq:multmom}, as we do for the Kerr-Newman metric in $\S 4$.

\section{Non-Einstein gravity}
Having introduced the Geroch-Hansen procedure for GR in $\S 2$B, we explore the parts that break down when the spacetime is non-Einstein.

\subsection{Definitions}

For $R_{\mu \nu} \neq 0$, the twist one-form $\boldsymbol{\omega}^{\dag}$ is not curl-free. We introduce a one-form $\boldsymbol{Q}$ such that
\begin{equation} \label{eq:gentwist22}
\boldsymbol{\omega} = \boldsymbol{\omega^{\dag}} + \boldsymbol{Q}  
\end{equation}
is curl-free. Then a scalar field $\omega$ exists with $\nabla_{\alpha} \omega = \omega_{\alpha}$, provided that $\boldsymbol{Q}$ satisfies
\begin{equation} \label{eq:qrelation}
d \boldsymbol{Q} = - \star \left[ \boldsymbol{\xi} \wedge \boldsymbol{R} \left(\boldsymbol{\xi}\right)\right] ,
\end{equation}
where $\star$ is the Hodge star operator and $\wedge$ is the wedge product (see equation 2.16 of \cite{heusler}). Equation \eqref{eq:qrelation} constitutes a first-order partial differential equation for the form $\boldsymbol{Q}$ \cite{brezis}, which may be solved  by defining a coordinate system (see Appendix C).  We explicitly construct $\boldsymbol{Q}$ for the case of the Kerr-Newman black hole in $\S 4$.


\subsection{Moments}
In this section we derive the equations for the gravitational potentials which define the multipole moments. We introduce scalar fields $\tau^{A}(\lambda,\omega)$, generalising the moment functions $\phi^{A}$ of Geroch and Hansen. The goal is to choose $\tau^{A}$ to satisfy equation \eqref{eq:poisson} for $R_{\mu \nu} \neq 0$, with $\tau^{A} = \phi^{A}$ for $R_{\mu \nu} = 0$. Making use of the chain rule $D_{j} \tau^{A} = D_{j} \lambda \frac {\partial \tau^{A}} {\partial \lambda} + D_{j} \omega \frac {\partial \tau^{A}} {\partial \omega}$,  we are able to write
\begin{equation} \label{eq:anotherpois}
\begin{aligned}
&(D^{m} D_{m} - \tfrac {1} {8} \mathscr{R})\tau^{A} \\
&= \frac {1} {\sqrt{h}} \partial_{i} \Big( \sqrt{h} h^{ij} \partial_{j} \tau^{A} \Big) - \tfrac {1} {8} \mathscr{R} \tau^{A} \\
&= \left( D^{m} D_{m} \lambda \right) \tau^{A}_{,\lambda}  + \left( D^{m} D_{m} \omega \right) \tau^{A}_{,\omega} + \left( D_{i} \lambda D^{i} \lambda \right) \tau^{A}_{,\lambda \lambda} \\
&\,\,\,\,\,\, + 2 \left( D^{j} \omega D_{j} \lambda \right) \tau^{A}_{,\lambda \omega} + \left( D^{i} \omega D_{i} \omega \right) \tau^{A}_{,\omega \omega} - \tfrac{1}{8}\mathscr{R}\tau^{A} ,
\end{aligned}
\end{equation}
where we have adopted the notation $f^{A}_{,\lambda \omega} = \frac {\partial^2 f^{A}} {\partial \lambda \partial \omega}$ and $\boldsymbol{h}$ forms the 3-metric as in \eqref{eq:papapet}. Making use of some identities derived in Appendix A [equations \eqref{eq:ric}--\eqref{eq:lambda}] we find that the vanishing of the Poisson equation
\begin{equation} \label{eq:nongrpois}
\left(D^{m} D_{m} - \tfrac {1} {8} \mathscr{R}\right)\tau^{A} = \tfrac{15}{16} \lambda^{-2} \kappa \tau^{A},
\end{equation}
is equivalent to
\begin{widetext}
\begin{equation} \label{eq:fieldons}
\begin{aligned}
0 =& D^{i} \lambda D_{i} \lambda \left(\tfrac {1} {2} \lambda^{-1} \tau^{A}_{,\lambda} + \tau^{A}_{,\lambda \lambda} - \tfrac{15}{16} \lambda^{-2} \tau^{A} \right) + D^{m} \omega D_{m} \omega \left( \tau^{A}_{,\omega \omega} - \lambda^{-1} \tau^{A}_{,\lambda} - \tfrac {3} {4} \lambda^{-2} \tau^{A} \right) + D^{m} \omega D_{m} \lambda \left( 2 \tau^{A}_{,\lambda \omega} + \tfrac {3} {2} \lambda^{-1} \tau^{A}_{,\omega} \right) \\
&+ D^{m} \omega Q_{m} \big( 2 \lambda^{-1} \tau^{A}_{,\lambda} - \tfrac {3} {8} \lambda^{-2} \tau^{A} \big) - \lambda^{-1} Q^{m} Q_{m} \left( \tau^{A}_{,\lambda} - \tfrac {3} {16} \lambda^{-1} \tau^{A} \right) + \left( D^{a} Q_{a} - \tfrac {3} {2} \lambda^{-1} Q_{m} D^{m} \lambda \right) \tau^{A}_{,\omega} - 2 R^{mn} \xi_{m} \xi_{n} \tau^{A}_{,\lambda} \\
&- \frac {R^{mn}} {8} \left( h_{mn} - \lambda^{-1} \xi_{m} \xi_{n} \right) \tau^{A}. \\
\end{aligned}
\end{equation}
\end{widetext}
For the mass potential, the above equation is to be solved subject to the Neumann conditions $\tau^{M}(\lambda=1,\omega) = \tfrac{\omega^2}{4}$ and $\tau_{,\lambda}^{M}(\lambda=1,\omega) = \tfrac{1}{2} - \tfrac{3}{16}\omega^2$. For the current potential, it is solved subject to the mixed Dirichlet $\tau^{J}(\lambda,\omega=0) = 0$ and Neumann conditions $\tau^{J}_{,\omega}(\lambda,\omega=0) = \tfrac{1}{2}$ \cite{hansen}. Adding constant terms to $\tau^{A}$ does not change the moment structure (because we have $\Omega|_{\Lambda} = 0$), so these boundary conditions are sufficient \cite{beigsim}. For the Einstein case $R_{\mu \nu} = 0$, equation \eqref{eq:fieldons} has solutions $\tau^{M} \rightarrow \phi^{M} = \tfrac{1}{4} \lambda^{-3/4} \left(\lambda^2 + \omega^2 - 1\right)$, and $\tau^{J} \rightarrow \phi^{J} = \tfrac{1}{2} \lambda^{-3/4} \omega$, which differ from Hansen's by a factor of $\lambda^{-1/4}$ due to the sign of the norm \eqref{eq:norm} (see section 7 of \cite{backdahlh}).

For any metric, the equations of motion for $\lambda$ and $\omega$ may be replaced by \eqref{eq:fieldons}, which reduces to the Laplace equation for $\boldsymbol{g} = \boldsymbol{\eta}$. There are also equations of motion for $h_{ij}$, which depend on the theory of gravity but do not affect the moments. As an example, we derive the field equations for $h_{ij}$ for $f(R)$ gravity in Appendix A.

We explicitly solve \eqref{eq:fieldons} in $\S4$ for two cases of `charged' black holes, which arise as vacuum solutions in an $f(R)$ theory of gravity: the Reissner-N{\"o}rdstrom and Kerr-Newman metrics. The moments for the modified gravity solution are given through \eqref{eq:multmom} with $\phi^{A}$ replaced by $\tau^{A}$. The process of computing $\boldsymbol{P}^{A}$ iteratively from \eqref{eq:multmom} after finding $\tau^{A}$ may or may not be a difficult task. Several authors \cite{backdahlh,gurl1,gurl2,fodor} have developed alternative methods for computing the moments rather than employing the definition \eqref{eq:multmom} directly, which are useful for complicated metrics.

\subsection{Static spacetimes}
For $\boldsymbol{\omega^{\dag}} = 0$ and $\boldsymbol{Q} = 0$, equation \eqref{eq:fieldons} gives
\begin{equation} \label{eq:staticcase}
\begin{aligned}
&D^{i} \lambda D_{i} \lambda \left( \tfrac{1}{2}\lambda^{-1}\tau^{A}_{,\lambda} + \tau^{A}_{,\lambda \lambda} - \tfrac{15}{16} \lambda^{-2} \tau^{A} \right) \\
&+ \tfrac{1}{8} R_{mn} \left( h^{mn} \tau^{A} - \lambda^{-1} \xi^{m} \xi^{n} \tau^{A} - 16 \xi^{m} \xi^{n} \tau^{A}_{,\lambda} \right) = 0.
\end{aligned}
\end{equation}
Furthermore, for $R_{\mu \nu} = 0$, \eqref{eq:staticcase} reduces to $(D_{i} \lambda \neq 0)$
\begin{equation} \label{eq:einsteincasestatic}
\tfrac {1} {2} \lambda^{-1} \tau^{A}_{,\lambda} + \tau^{A}_{,\lambda \lambda} - \tfrac{15}{16} \lambda^{-2} \tau^{A} = 0,
\end{equation}
which has solutions $\tau^{M} \rightarrow \phi^{M} =  \tfrac{1}{4}\lambda^{-3/4} \left(-1 + \lambda^2\right)$ and $\tau^{J} = 0$, where the dependence on $\lambda$ differs from Geroch's result \cite{gerochcurvedmult} due to $\kappa \neq 0$. Taking $\kappa = 0$, the appropriately modified version of equation \eqref{eq:einsteincasestatic} reads
\begin{equation}
\tfrac {1} {2} \lambda^{-1} \tau^{A}_{,\lambda} + \tau^{A}_{,\lambda \lambda} = 0,
\end{equation}
which has the exact Geroch solutions $\tau^{M} \rightarrow \phi^{M} =  -1 + \sqrt{\lambda}$ and $\tau^{J} = 0$ \cite{gerochcurvedmult,backdahlh}.

\subsection{Axisymmetric spacetimes}
In spacetimes that are axisymmetric, the tensor moments \eqref{eq:multmom} reduce to a set of scalar moments \cite{hansen}. Axisymmetry guarantees the existence of a spacelike Killing vector $\boldsymbol{\zeta}$ in $S$, i.e. there is a second field satisfying Killing's equation \eqref{eq:killing}. The conformal factor $\Omega$ can be chosen, without loss of generality, such that the axis vector $z^{i} = 2 \epsilon^{ijk} D_{j} \zeta_{k}$ satisfies
\begin{equation}
\tilde{z}^{k} \tilde{z}_{k}|_{\Lambda} = 1,
\end{equation}
under the conformal transformation $\boldsymbol{z} \rightarrow \tilde{\boldsymbol{z}}$; see equations (3.2)--(3.4) in Hansen \cite{hansen}. Hence, at $\Lambda$, $\boldsymbol{\zeta}$ defines rotations of tensors, under which the multipole moments are preserved \cite{hansen,papsot}. Hansen showed that the only tensors at $\Lambda$ invariant under the action of $\tilde{\boldsymbol{z}}$ are products of the metric and $\tilde{\boldsymbol{z}}$ itself \cite{hansen}. As a result, the $2^{s}$ moment $P^{A}_{a_{1}\cdots a_{s}}$ is necessarily a multiple of $\mathscr{C} \left[ \tilde{z}_{i_{1}} \cdots \tilde{z}_{i_{a_{s}}} \right]|_{\Lambda}$. Therefore, the only non-zero components of $\boldsymbol{P}^{A}$ are \cite{papsot,backdahlh,fodor}
\begin{equation} \label{eq:axisym}
P^{A}_{s} = \frac {1} {s!} P^{A}_{a_{1} \cdots a_{s}} z^{i_{1}} \cdots z^{i_{a_{s}}} \Big|_{\Lambda}.
\end{equation}

\section{`Charged' black holes}
In this section we solve equation \eqref{eq:fieldons} for the Reissner-N{\"o}rdstrom and Kerr-Newman metrics. Although originally obtained as charged black holes in GR, these black holes can also arise as vacuum solutions in modified theories of gravity, such as the $f(R)$ theory, explaining the quotation marks in the section heading. Our analysis follows various authors \cite{johpsa,glampbab,vigeland} in evaluating the multipole moments associated with a non-Einstein spacetime. The main distinction here is that the generalized Geroch-Hansen procedure allows us to directly compare the moments for exact solutions in modified gravitational theories with their Einstein counterparts at all orders.

\subsection{Reissner-N{\"o}rdstrom metric}
A static, spherically symmetric metric on $M$ may be written in Boyer-Lindquist $\left(t,r,\theta,\phi\right)$ coordinates as (see e.g. \cite{wald})
\begin{equation} \label{eq:genstatmet}
ds_{M}^2 = A(r) dt^2 - B(r) dr^2 - r^2 d \Omega^2  .
\end{equation}
Static solutions have $\boldsymbol{\omega^{\dag}} = 0$, and $\lambda = A(r)$. From \eqref{eq:papapet}, the line element on $S$ is
\begin{equation} \label{eq:line1}
d \sigma^2_{S} = B(r) A(r) dr^2 + r^2 A(r) d \Omega^2 .
\end{equation}
The Einstein-Maxwell field equations read
\begin{equation} \label{eq:maxwell}
R^{\mu \nu} - \frac {R} {2} g^{\mu \nu} = \frac {1} {\mu_{0}} \left( F^{\mu \alpha} F^{\nu}_{\alpha} - \frac {1} {4} g^{\mu \nu} F^{\alpha \beta} F_{\alpha \beta} \right),
\end{equation}
where $F^{\alpha \beta}$ is the Faraday tensor (e.g. \cite{heusler}). Choosing a point source to generate $\boldsymbol{F}$ yields the Reissner-N{\"o}rdstrom solution (e.g. \cite{mtw})
\begin{equation} \label{eq:reissner}
A(r) = 1 - \frac {2M} {r} + \frac {\rho} {r^2} = B(r)^{-1} .
\end{equation}
The parameter $\rho$, for the Einstein-Maxwell case, is defined as the square of the total electromagnetic charge \cite{mtw},
\begin{equation} \label{eq:komar}
\rho = q^2.
\end{equation}
We avoid writing $\rho$ as a square, because $\rho$ can be interpreted as a negative number in modified gravity. To the authors' knowledge, there is no neat physical interpretation for $\rho$ presently available in an $f(R)$ gravity where there is zero electromagnetic field. In the vacuum theory it appears as an integration constant, and seems to have the effect of mass renormalization (see $\S 4$ B).

The Geroch-Hansen moments in GR are not able to describe the multipolar decomposition of this solution\footnote{However one can define separate moments for the Schwarzschild component and electromagnetic 4-potential $\boldsymbol{A}$; see \cite{hoensel,sotetal,aposot,sotapo}.} since $R_{\mu \nu} \neq 0$. However, the Reissner-N{\"o}rdstrom metric is in fact a solution to any vacuum $f(R)$ gravity with $f(0) = f'(0) = 0$, such as the $f(R) = R^{1+\delta}$ models studied by Clifton and Barrow \cite{clifbar}. In this paper we consider models with $\delta > 0$. Clifton and Barrow found that perihelion precession observations of Mercury place the tight bounds $ 0 \leq \delta < 7.2 \times 10^{-19}$ \cite{clifbar}.

 The field equations for $f(R)$ gravity read (e.g. \cite{felice})
\begin{equation} \label{eq:fofrgrav}
f'(R) R_{\mu \nu} - \frac {1} {2} f(R) g_{\mu \nu} + (g_{\mu \nu} \square - \nabla_{\mu} \nabla_{\nu} ) f'(R) = 0  .
\end{equation}
Noting that the Reissner-N{\"o}rdstrom spacetime has $R=0$, setting $f(0) = f'(0) = 0$ automatically guarantees that \eqref{eq:genstatmet} with \eqref{eq:reissner} is a solution to \eqref{eq:fofrgrav}, since $\nabla_{\mu} f'(R) = R_{;\mu} f''(R)$ and $R$ is constant.

Using our expression for the 3-metric \eqref{eq:line1}, equation \eqref{eq:staticcase} for the Reissner-N{\"o}rdstrom solution becomes
\begin{widetext}
\begin{equation}
\begin{aligned}
0 =& D^{i} \lambda D_{i} \lambda \left( \frac {1} {2} \lambda^{-1} \tau^{M}_{,\lambda} + \tau^{M}_{,\lambda \lambda} - \tfrac{15}{16} \lambda^{-2} \tau^{M} \right) \\
&- \frac {\left(\lambda -1\right)^{4} \rho \Big\{ 8 M^2 p(\lambda)^{2} + (\lambda - 1)^{2} \rho^{2} + 4 M \left[p(\lambda)^{2} + M^2\right] p(\lambda) \Big\}} {8 \lambda \left[ M + p(\lambda) \right]^{8} } \left[ \left( \lambda -1 \right) \tau^{M} + 16 \lambda^2 \tau^{M}_{,\lambda} \right] = 0,
\end{aligned}
\end{equation}
\end{widetext}
with $p(\lambda) = \left[ M^2 + (\lambda - 1) \rho \right]^{1/2}$. We have $\tau^{J} = 0$ by the boundary conditions, as expected physically. Explicitly taking the covariant derivatives of $\lambda$, i.e. \nolinebreak $D_{i} \lambda = D_{i} A(r) = \delta^{r}_{i} A'(r) = \delta^{r}_{i} A'( A^{-1}(\lambda))$, we find
\begin{equation} \label{eq:rnpde}
\begin{aligned}
0 =& \left[\left(\lambda - 1 \right) \rho + 30 \lambda^{-1} p(\lambda)^{2} \right] \tau^{M} - 16 \left[ p(\lambda)^2 - \lambda^2 \rho \right] \tau^{M}_{,\lambda} \\
&- 32 \lambda p(\lambda)^2 \tau^{M}_{,\lambda \lambda}.
\end{aligned}
\end{equation}
Equation \eqref{eq:rnpde} with boundary conditions $\tau^{M}(\lambda = 1) = 0$, $\tau^{M}_{,\lambda}(\lambda = 1) = 1/2$ has a lengthy solution involving confluent hypergeometric functions \cite{kamke}, which is plotted in Figure 1 for a variety of values of $\rho$. 

\begin{figure*}
\includegraphics[width=\textwidth]{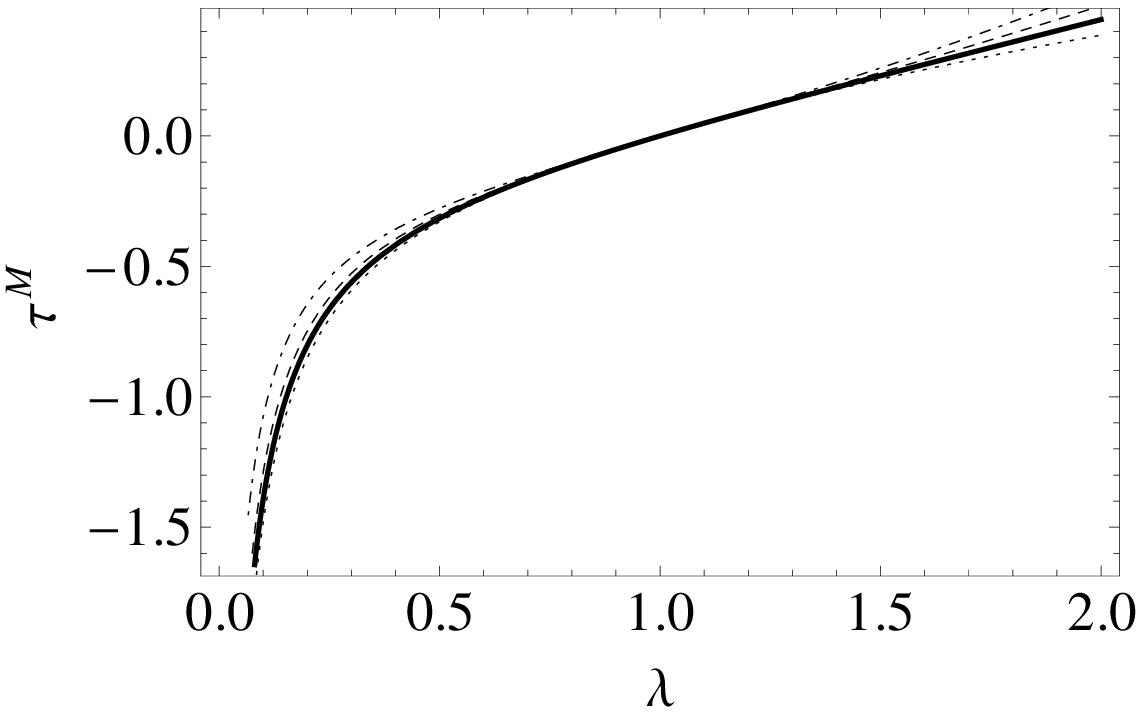}
\justifying
FIG. 1: Solutions $\tau^{M}(\lambda)$ to the Poisson equation \eqref{eq:rnpde} with $\rho = -0.4M^2$ (dotted), $\rho = 0.0$ (solid), $\rho = 0.4 M^2$ (dashed), and $\rho = 0.99M^2$ (dash-dotted).
\end{figure*}

We now compute the moments following the recipe in $\S 2$ and $\S 3$. Following Hansen \cite{hansen}, we adopt a new coordinate system $\{t,\bar{R},\theta,\phi\}$ which simplifies the calculation of $\Omega$ considerably, where the new radial coordinate $\bar{R}$ is defined via
\begin{equation} \label{eq:radialco}
\bar{R} = \exp \left[ \int dr \sqrt{\frac {h_{rr}} {h_{\theta \theta}}} \right] .
\end{equation}
Upon inversion for the Reissner-N\"ordstrom spacetime, we find
\begin{equation} \label{eq:conformal12}
r = \frac {1} {\bar{R}} \left( 1 + M \bar{R} + \frac {1} {4} \left(M^2 - \rho\right) \bar{R}^2 \right) .
\end{equation}
Inspection of the resulting line element shows that a suitable $\Omega$ can be taken as
\begin{equation} \label{eq:conformal23}
\Omega = \frac {\bar{R}^2} {1- \frac {1} {4} (M^2 - \rho )\bar{R}^2} ,
\end{equation}
to yield
\begin{equation} \label{eq:conlinel}
\tilde{d \sigma}_{\tilde{S}}^2 = d \bar{R}^2 + \bar{R}^2 d \theta^2 + \bar{R}^2 \sin^2\theta d \phi^2 ,
\end{equation}
satisfying $\boldsymbol{\tilde{h}} = \Omega^{2} \boldsymbol{h}$. These choices reduce to those used by Geroch and Hansen \cite{gerochcurvedmult,hansen} when $\rho \rightarrow 0$. Upon substituting $\tilde{\tau}^{M} = \Omega^{-1/2} \tau^{M}$ into \eqref{eq:axisym}, we find that the only non-zero multipole moment for the $f(R)$ Reissner-N{\"o}rdstrom spacetime is
\begin{equation} \label{eq:rnmult12}
M_{0} = - \left(M^2 - \rho\right)^{1/2} .
\end{equation}
Equation \eqref{eq:rnmult12} agrees with Geroch's result $M_{0} = - M$ as a special case for the Schwarzschild solution ($\rho \rightarrow 0$).

\subsection{Kerr-Newman solution}
The Kerr-Newman metric in Boyer-Lindquist-type $\left(t,r,\theta,\phi\right)$ coordinates takes the form (e.g. \cite{chandbh})
\begin{equation*} 
g_{tt} = 1 + \frac {\rho - 2 M r} {r^2 + a^2 \cos^2\theta}  ,
\end{equation*}
\begin{equation*}
g_{rr} = - \frac {r^2 + a^2 \cos^2\theta} {a^2 - 2 M r + r^2 + \rho} ,
\end{equation*}
\begin{equation} \label{eq:kernew1}
g_{t \phi} = \frac {a (2 M r - \rho) \sin^2\theta} {r^2 + a^2 \cos^2\theta} ,
\end{equation}
\begin{equation*}
g_{\theta \theta} = - r^2 - a^2 \cos^2\theta ,
\end{equation*}
\begin{equation*} 
g_{\phi \phi} = -\left( r^2 + a^2 - \frac {a^2(\rho - 2 M r) \sin^2\theta} {r^2 + a^2 \cos^2\theta} \right) \sin^2\theta .
\end{equation*}
The parameter $a = J/M$ measures the spin, and $\rho$ is once again the total charge as in \eqref{eq:komar} for the Einstein-Maxwell theory (but not in modified gravity).

The Kerr-Newman metric, which has $R = 0$, is a vacuum solution to any $f(R)$ theory\footnote{This solution also appears as a vacuum solution in the Randall-Sundrum braneworld class of theories \cite{aliev}.} with $f(0) = f'(0) = 0$ by the same argument as in $\S 4$A. The Ricci tensor has a non-vanishing off-diagonal component
\begin{equation} \label{eq:rictensortp}
R_{t}^{\phi} = \frac {2 a \rho} {(r^2 + a^2 \cos^2\theta)^{3}} ,
\end{equation}
so the generalised twist must be constructed. Solving equation \eqref{eq:qrelation} using the equivalent system \eqref{eq:qrelation4}-\eqref{eq:qrelation6} we find $\boldsymbol{Q}$ has non-zero components
\begin{equation} \label{eq:kerrq}
Q_{r} = - \frac {2 a \rho \cos\theta} {\left( r^2 + a^2 \cos^2\theta \right)^{2}},
\end{equation}
and
\begin{equation} \label{eq:kerrq2}
Q_{\theta} = - \frac {2 a r \rho \sin \theta} {\left( r^2 + a^2 \cos^2\theta \right)^{2}},
\end{equation}
and hence
\begin{equation} \label{eq:kerromega}
\omega = \frac {2 a M \cos \theta} {r^2 + a^2 \cos\theta},
\end{equation}
from \eqref{eq:curl} and \eqref{eq:gentwist22}. Equation \eqref{eq:fieldons} may now be written down in its entirety. We do not record the result here as it is lengthy. We use the relation $\lambda = g_{tt}$ and equation \eqref{eq:kerromega} to express the coordinates $r$ and $\theta$ as well the components of $\boldsymbol{Q}$ in terms of $r = r(\lambda, \omega)$, $\theta = \theta(\lambda,\omega)$, and $Q_{\alpha} = Q_{\alpha}(\lambda,\omega)$. All terms involving the Ricci tensor may then be expressed in terms of $\lambda$ and $\omega$.

Defining a new radial coordinate $\bar{R}$ through the implicit relationship
\begin{equation}
r = \frac {1} {\bar{R}} \left[ 1 + M \bar{R} + \frac {1} {4} \left(M^2 - a^2 - \rho\right) \bar{R}^2 \right] ,
\end{equation}
we find that
\begin{equation}
\Omega = \frac {\bar{R}^2} { \Big\{\left[1 - \frac {1} {4} \left(M^2 - a^2 - \rho\right) \bar{R}^2\right]^{2} - a^2 \bar{R}^2 \sin^2\theta \Big\}^{1/2}} ,
\end{equation}
yields the desired form of the line element
\begin{equation} \label{eq:someline}
\begin{aligned}
\tilde{d \sigma^{2}_{S}} = &d \bar{R}^2 + \bar{R}^2 d \theta^2 + \\
&\frac {\bar{R}^2 \sin^2\theta d \phi^2} {1 - a^2 \bar{R}^2 \sin^2\theta [1 - \frac {1} {4} (M^2 - a^2 - \rho) \bar{R}^2]^{-2} } ,
\end{aligned}
\end{equation}
with $\boldsymbol{\tilde{h}} = \Omega^2 \boldsymbol{h}$. A quick check shows that \eqref{eq:someline} does satisfy the constraints at the point $\Lambda$ $(\bar{R} = 0)$.

Solving \eqref{eq:fieldons} and iteratively evaluating the moments $\boldsymbol{P}^{A}|_{\Lambda}$ through equation \eqref{eq:axisym} we obtain the multipole moments for the Kerr-Newman solution in $f(R)$ gravity:
\begin{equation} \label{eq:kerrnewmanmult1}
M_{2s} = (-1)^{s+1} \left(M^2 - \rho \right)^{1/2} a^{2 s},
\end{equation}
\begin{equation} \label{eq:kerrnewmanmult2}
M_{2s +1} = 0,
\end{equation}
\begin{equation} \label{eq:kerrnewmanmult3}
J_{2s+1} = (-1)^{s} \left(M^2 - \rho\right)^{1/2} a^{2 s +1},
\end{equation}
\begin{equation} \label{eq:kerrnewmanmult4}
J_{2s} = 0.
\end{equation}
Again, these results agree with those found by Hansen \cite{hansen} for the Kerr solution in the limit $\rho \rightarrow 0$. It is worth pointing out that the moments may each be written in terms of $M_{0}$ and $J_{1}$ (instead of $M, \rho$, and $a$) to recover the same expressions as in Kerr, i.e.
\begin{equation} \label{eq:inspect1}
M_{2s} = (-1)^{s} M_{0} a^{2 s},
\end{equation}
and
\begin{equation} \label{eq:inspect2}
J_{2s+1} = (-1)^{s} J_{1} a^{2 s}.
\end{equation}
Therefore, in this sense, the (Kerr) no-hair relation is not violated. Moreover, this allows one to interpret $\rho$ as an $f(R)$ renormalization of the mass term in the Kerr metric; if the Kerr mass $M$ is replaced by the $f(R)$ term $\sqrt{M^2 - \rho}$, then the Kerr moments are indistinguishable from the $f(R)$ Kerr-Newman moments.

\section{QPOs in microquasars}
To illustrate how the above results can be applied to practical problems, we consider X-ray observations of QPOs in accreting compact objects, especially microquasars \cite{mcclintock,johpsa}. There are several models in the literature for the origin of QPOs (e.g. \cite{stefanov}). One possibility is that QPOs arise from normal modes of oscillation trapped due to Lense-Thirring precession in the accretion disks surrounding black holes \cite{wagoner,wagonerdisk}. We explore two simplified models along these lines: the diskoseismology model (e.g. \cite{perez}) in $\S 5$A and the kinematic resonance model (e.g. \cite{kluz}) in $\S 5$B. In the diskoseismology model, QPO frequencies are attributed to the lowest order gravity ($g$-) and corrugation ($c$-) modes \cite{wagoner12}. In the kinematic resonance model, the QPOs arise from resonances between the radial epicyclic frequencies of disk particles and the corresponding Keplerian frequencies \cite{kluz}. Assuming particular resonances for the objects GRS $1915$+$105$ and GRO J$1655$-$40$, we explore the range of spin $a$ and `charge' $\rho$ values compatible with the QPO models above and independent mass measurements. We stick to using the parameter $\rho$ throughout this section, although we can easily solve for (say) the mass quadrupole moment $M_{2}$ defined in \eqref{eq:kerrnewmanmult1} from
\begin{equation} \label{eq:quadrho}
\rho = M^2 -  M_{2}^{2}/a^{4}.
\end{equation}
There are large systematic uncertainties in determining black hole spins from QPOs, because there are many viable models through which the data can be interpreted. The purpose of this section is to point out that there may also be uncertainties in the gravitational theory, which compound those that enter through QPO modelling.

\subsection{Diskoseismology model}
For a massive particle falling freely in a stationary, axisymmetric metric with four-momentum
\begin{equation} \label{eq:fourmom}
p^{\alpha} = \mu \frac {d x^{\alpha}} {d \tau},
\end{equation}
we have three conserved quantities: rest mass $\mu$ (which we set to 1 without loss of generality), energy $E = -p_{t}$, and axial angular momentum $L_{z} = p_{\phi}$ \cite{chandbh}. Raising the indices on these momenta we obtain the relations \cite{glampbab}
\begin{equation} \label{eq:pt}
p^{t} = - \frac {g_{\phi \phi} E + g_{t \phi} L_{z}} {g_{tt} g_{\phi \phi} - g_{t\phi}^2} ,
\end{equation}
\begin{equation} \label{eq:pphi}
p^{\phi} =  \frac {g_{t \phi} E + g_{t t} L_{z}} {g_{tt} g_{\phi \phi} - g_{t\phi}^2} .
\end{equation}
If we specialize to motion in the equatorial plane $(p^{\theta} = 0)$, appropriate for a thin, unwarped accretion disk, the remaining independent component of the geodesic equations turns into
\begin{equation} \label{eq:geodesic}
\frac {1} {2} g_{rr} \left( p^{r} \right)^{2} = \frac {1} {2} \left[ - g_{tt} \left( p^{t} \right)^2  - 2g_{t\phi}  p^{t} p^{\phi}  - g_{\phi\phi} \left( p^{\phi} \right)^2 -1 \right] . 
\end{equation}
The right-hand side of \eqref{eq:geodesic}, written $V_{\text{eff}}$, is called the effective potential. Circular orbits are then characterised by the additional constraints $p^{r} = 0$ and $\dfrac {d p^{r}} {d r}= 0$ \cite{chandbh,johpsa}. Specialising to this case, we have from equation \eqref{eq:geodesic}
\begin{equation} \label{eq:veff1}
V_{\text{eff}} = 0 ,
\end{equation}
and
\begin{equation} \label{eq:veff2}
\frac {d} {d r} \left( \frac {V_{\text{eff}}} {g_{rr}} \right) = 0 .
\end{equation}
Using the momenta \eqref{eq:pt} and \eqref{eq:pphi}, solving equations \eqref{eq:veff1} and \eqref{eq:veff2} leads to expressions for the energy
\begin{equation} \label{eq:energy}
E = \frac {-2Mr + r^2 \pm a \left(Mr - \rho\right)^{1/2} - a \rho} {r \sqrt{ -3 M r + r^2 \pm 2 a\left(Mr - \rho\right)^{1/2} + 2 \rho}} ,
\end{equation}
and axial angular momentum
\begin{equation} \label{eq:angul}
L_{z} = \pm \frac {\left[ a^2 + r^2  \mp 2 a \left( M r -\rho \right)^{1/2} \right] \left(M r - \rho \right)^{1/2} \mp a \rho} {r \sqrt{-3 M r + r^2 \pm 2 a\left( M r - \rho\right)^{1/2} + 2 \rho}},
\end{equation}
where the upper sign is taken for prograde orbits and the lower sign for retrograde orbits. Restoring dimensional factors of $G$ and $c$ and using expressions \eqref{eq:energy} and \eqref{eq:angul} we define the Keplerian frequency $\Omega_{\phi}$,
\begin{equation} \label{eq:kepler}
\Omega_{\phi} = - \frac {c^3} {G} \frac {g_{t\phi} E + g_{tt} L_{z}} {g_{\phi\phi} E + g_{t \phi} L_{z}},
\end{equation}
the radial epicyclic frequency $\kappa_{r}$ given by \cite{johpsa}
\begin{equation} \label{eq:radialepi}
\kappa_{r}^2 = - \left( \frac {c^3} {G}\right)^{2} \frac {\partial^2} {\partial r^2} \left[ \frac {V_{\text{eff}}} {g_{rr} \left( p^{t} \right)^{2}} \right]  ,
\end{equation} 
and the angular epicyclic frequency $\Omega_{\theta}$ given by \cite{johpsa}
\begin{equation} \label{eq:angularepi}
\Omega_{\theta}^2 = -  \left( \frac {c^3} {G} \right)^{2} \frac {\partial^2} {\partial \theta^2} \left[ \frac {V_{\text{eff}}} {g_{\theta\theta} \left( p^{t} \right)^{2}} \right] ,
\end{equation}
evaluated in the equatorial plane $\theta = \pi/2$. The expressions for these quantities as functions of $r$, $E$, and $L_{z}$ are lengthy but can be evaluated easily using the metric components written down in $\S 4$B. 

Following previous authors \cite{wagoner,wagonerdisk}, we assume that the fundamental $g$-mode frequency is given by the maximum value of $\kappa_{r} / 2\pi$ with respect to $r$, while the fundamental $c$-mode frequency is given by the Lense-Thirring frequency $\Omega_{\text{LT}} = \Omega_{\phi} - \Omega_{\theta}$ (e.g. \cite{chandbh}) evaluated at the innermost stable circular orbit (ISCO) \cite{johpsa}. The radius of the ISCO, $r_{\text{ISCO}}$, for the Kerr-Newman metric is given as the smallest (real) solution to the equation,
\begin{equation}
\begin{split}
0 =& M \left( 6 M -r_{\text{ISCO}} \right) r_{\text{ISCO}}^2 + a^2 \left( 3 M r_{\text{ISCO}} - 4 \rho \right) \\
&\pm 8 a \left(M r_{\text{ISCO}} - \rho \right)^{3/2} - 9 M r_{\text{ISCO}} \rho + 4 \rho^2.
\end{split}
\end{equation}

In Figure 2 we plot the resulting $g$- and $c$-mode frequencies for the Kerr-Newman metric as a function of spin for a range of values of $\rho$. As $\rho$ increases in the positive direction, the frequency of the relevant $g$-mode can become very large $(\gtrsim 0.3 \text{kHz})$, when $a/M \sim 0.9$ for a $10 M_{\odot}$ object, as can the frequency of the $c$-mode $(\gtrsim \text{kHz})$. In GR, the $g$-mode frequency is bounded above by $ 248.3 (M/10M_{\odot}) \text{Hz}$ for the extreme case $a=M$. Allowing for positive values of $\rho$ in $f(R)$ gravity permits the frequency to exceed this bound. Furthermore, the $c$-mode frequency is undefined in the presence of a naked singularity, which forms for the Kerr-Newman solution when $a > \sqrt{3} M / \sqrt{5}$ for $\rho = 0.4 M^2$ as indicated by the spike in the dashed curve. We see that for $0.1 \lesssim a/M \lesssim 0.4 $ the $g$-mode frequencies predicted for $\rho = 0.4 M^2$ and $\rho = 0$ are almost indistinguishable.

In Figure 3, we plot deviations in the $g$- and $c$-mode frequencies from the Kerr value ($\rho = 0$) for two different quasar systems GRS $1915$+$105$ and GRO J$1655$-$40$. To demonstrate, we take the Middelton et al. values \cite{middle} for the spin of GRS $1915$+$105$ $(a = 0.998 M)$ and the Motta et al. values \cite{motta} for the spin of GRO J$1655$-$40$ $(a = 0.29 M)$. These values were obtained by fitting continuum and precession models to the data \cite{bambi} and are compatible with our estimates computed in $\S 5$B. We see that increasing $\rho$ in the positive direction increases the frequencies of the $g$- and $c$-modes. For $a = 0.998 M$ in GRS $1915$+$105$, the model breaks down for $\rho \approx 10^{-2} M^2$ (naked singularity), as indicated by the spike in the $c$-mode frequency.

\begin{figure*}

\begin{subfigure}[b]{0.495\textwidth}
\includegraphics[width=\textwidth]{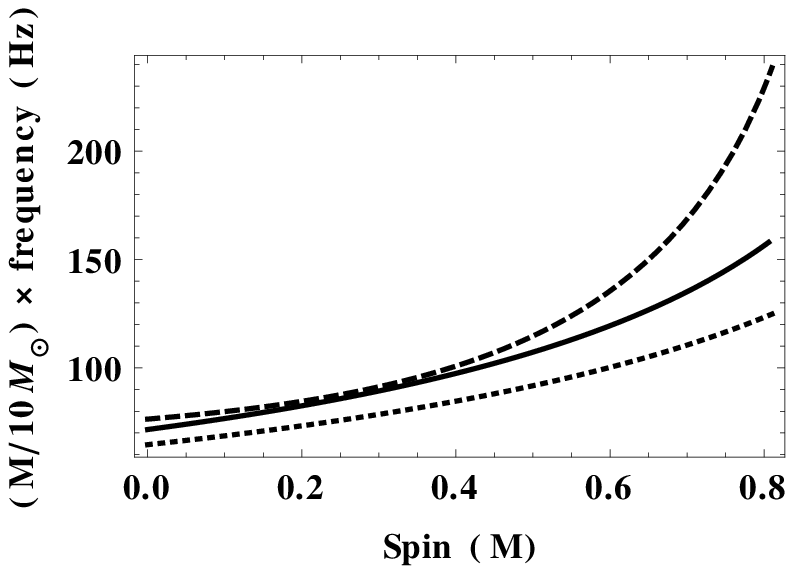}
\end{subfigure}
\begin{subfigure}[b]{0.495\textwidth}
\includegraphics[width=\textwidth]{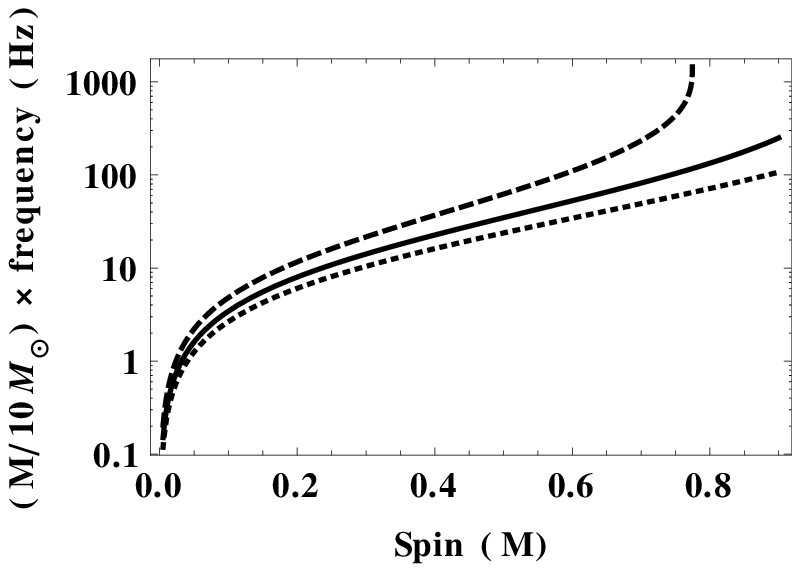}
\end{subfigure}
\justifying
FIG. 2: Fundamental $g$-mode (left panel) and $c$-mode (right panel) frequencies as functions of spin with $\rho = 0$ (solid), $\rho = 0.4 M^2$ (dashed) and $\rho = -0.4 M^2$ (dotted).
\end{figure*}

\begin{figure*}
\begin{subfigure}[b]{0.495\textwidth}
\includegraphics[width=\textwidth]{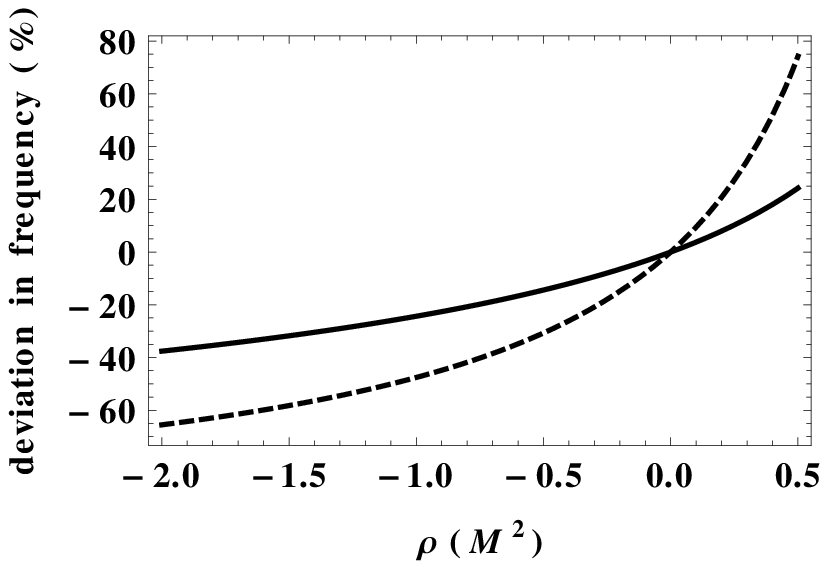}
\end{subfigure}
\begin{subfigure}[b]{0.495\textwidth}
\includegraphics[width=\textwidth]{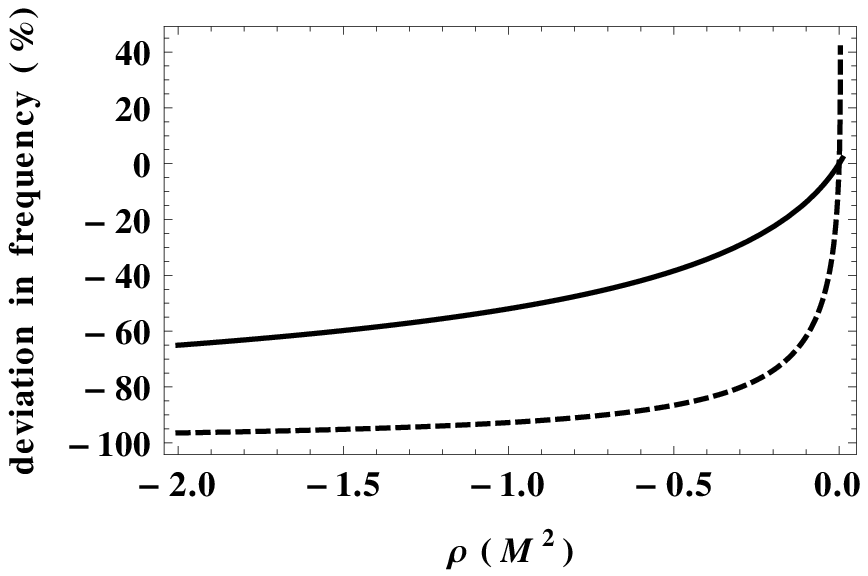}
\end{subfigure}
\justifying
FIG. 3: Deviations in predicted $g$-mode (solid curve) and $c$-mode (dashed curve) frequencies from the Kerr frequencies $(\rho = 0)$ as a function of $\rho$ for GRO J$1655$-$40$ (left panel) with spin parameter $a = 0.29 M$ and GRS $1915$+$105$ (right panel) with spin parameter $a = 0.998 M$.
\end{figure*}

\subsection{Kinematic resonance model}
In the kinematic resonance model, the measured QPO frequencies are identified with the Kepler frequency $\Omega_{\phi}$ and the radial epicyclic frequency $\kappa_{r}$, whose ratio is a simple fraction in resonance. Here we determine combinations of $a$ and $\rho$ which are consistent with the measured frequencies. 

QPO pairs are typically observed to have frequency ratios of $3/2$ or $1/2$ \cite{remillard}. Assuming these ratios are universal, several authors have estimated the spins of black holes assuming a Kerr description (for e.g. GRS $1915$+$105$ \cite{mcclintock} and others \cite{kluz,kluz2}). We show here that if the object is actually a Kerr-Newman black hole in $f(R)$ gravity, different spin estimates arise. If, in addition, one has an independent measure of the mass of the black hole, interesting bounds follow on $a$ and $\rho$. As $\rho$ can be swapped for the mass quadrupole moment (or any higher order moment) of the Kerr-Newman metric, we can also test whether or not the no-hair theorem is violated in GR, if $\rho = 0$ does not allow for any value of $a$ to fit the data. One needs an independent measurement of the spin to then also test whether or not any $f(R)$ no-hair relations are violated.

Independent measurements of the black hole mass in the systems GRS $1915$+$105$ $(M = 14.0 \pm 4.4 M_{\odot})$ and GRO J$1665$-$40$ $(M = 5.31 \pm 0.07 M_{\odot})$ have been made using  optical ray tracing \cite{harlaftis,harl2} and X-ray timing techniques \cite{motta}. Given a QPO pair at a given resonance with frequencies $q_{1}$ and $q_{2}$ we determine valid combinations of $a$ and $\rho$ as follows. We solve for the radius $r_{\text{res}}$ such that $\kappa_{r}(r_{\text{res}}) = q_{1}$ and determine values of $a$ and $\rho$ which yield $\Omega_{\phi}(r_{\text{res}}) = q_{2}$. We obtain a band of possible values owing to the error bars placed on the measured mass.


Using Doppler spectroscopy, Greiner et al. \cite{harl2} estimated the mass of GRS $1915$+$105$ via Kepler's Third Law with an orbital separation of $(108 \pm 4) R_{\odot}$, (i.e., separation $\gg r_{\text{Horizon}}$) from data taken at the Antu European South Observatory Very Large Telescope. The inferred mass is the same for both the Kerr and Kerr-Newman metrics, which have the same far-field (Newtonian) limit. In contrast, the more accurate mass estimate obtained for GRO J$1665$-$40$ through X-ray timing assumes a Kerr metric to analyse epicyclic frequencies \cite{motta}. In principle, this approach is incompatible with our analysis, since these frequencies, and hence the mass estimate, are computed from the metric components \eqref{eq:papapet} [see equations \eqref{eq:kepler}--\eqref{eq:angularepi}]. Independent optical measurements of the mass of GRO J$1665$-$40$, e.g. by Beer and Podsiadlowski \cite{beer} who found $M = 5.4 \pm 0.3 M_{\odot}$ from Kepler's Third Law, avoid this problem while remaining consistent with Motta et al. \cite{motta}. Our aim is to demonstrate that, even with tight error bars on the mass, as achieved by \cite{motta}, a large range of spins can accommodate QPO observations, a conclusion which is unaltered (but is less evident) if one uses the self-consistent values of the mass from Beer and Podsiadlowski \cite{beer}. As better data become available in the future, it would ultimately be worth re-calculating the measured mass based on the assumption of a central Kerr-Newman object following the rigorous statistical approach of Motta et al. \cite{motta} for an entirely self-consistent analysis, but this lies outside the scope of the present paper.


M{\'e}ndez et al. \cite{mendez} reported pairs of QPOs from GRS $1915$+$105$ with frequencies $35$ Hz and $67$ Hz, and from GRO J$1655$-$40$ with frequencies $300$ Hz and $450$ Hz. Clearly the QPOs come from $1$$:$$2$ and $2$$:$$3$ resonances respectively. For the rough estimates below we assume that these are exact values. The measured error bars are quite tight: $35.1 \pm 0.4$ Hz and $67.9 \pm 0.1$ Hz for GRS $1915$+$105$, and $289 \pm 4$ Hz and $446 \pm 2$ Hz for GRO J$1655$-$40$ respectively.

In Figure 4 we plot the possible parameter space for the object GRO J$1655$-$40$ with error bars flowing from $M$ as given by \cite{motta}. The $\sim 1\%$ error on $M$ results in a thin band, with $0 \leq a/M \leq 1$ and $-2.0 \lesssim \rho/M^2 \lesssim 0.5$. As $\rho$ increases in the positive direction, the required spin parameter decreases. This is in contrast to the behaviour of the quadrupole moment \eqref{eq:kerrnewmanmult1}, which decreases when either $\rho$ is positive and increasing, or $a$ decreases. Compatibility with the measured QPO frequencies then, for this model, demands that the quadrupole moment \emph{decreases} if $\rho$ is positive and increasing, or \emph{increases} if $\rho$ is negative and decreasing. For $\rho \gtrsim 0.35 M^2$ the lower end of the measured mass $(M \lesssim 5.25 M_{\odot})$ requires $a < 0$.

In Figure 5 we repeat the analysis for the object GRS $1915$+$105$ with error bars on the mass as given in \cite{harlaftis}. We obtain a large range of allowed combinations for $a$ and $\rho$ given the relatively poorly constrained mass measurement of $M = 14.0 \pm 4.4 M_{\odot}$. As $\rho$ increases from $0$ to $0.25 M^2$, the predicted range of $a$ drops from $0.2 \lesssim a/M \lesssim 0.999$ to $0.06 \lesssim a/M \lesssim 0.83$, meaning that GRS $1915$+$105$ may not be a near-extreme Kerr black hole as previously suspected \cite{middle}. In contrast to the GRO J$1655$-$40$ case, we find that for $\rho \gtrsim 0.25 M^2$ the top end of the measured mass range $(M \gtrsim 17 M_{\odot})$ does not allow for a solution, because the required value of $a$ exceeds the naked singularity threshold $a > \sqrt{3} / 2 M$.

We close this section with a suggested experiment to determine all three values $M$, $a$, and $\rho$. In reality, the quasar system may not be in equilibrium, and will be subject to perturbations, such as fluctuations in the accretion flow \cite{epois,yunes3}. These perturbations will distort the mass quadrupole moment, resulting in the emission of gravitational waves \cite{wald,ryan}, whose luminosities will be sensitive to the governing theory of gravitation \cite{delcap,bara}. Given a combination of gravitational wave detections (from e.g. LIGO \cite{ligo1}) and QPO measurements one could uniquely determine $a$, $M$, and $\rho$ (e.g. \cite{apoluk}). Some perturbation theory has been developed for $f(R)$ gravity (e.g. \cite{capbeyond,felice}) and would be useful in this context. Furthermore, $M$, $a$, and $\rho$ may also be determined independently by measuring multipole moments of a central black hole from gravitational wave observations of an extreme-mass ratio inspiral, as pioneered by Ryan \cite{ryan,ryan2}.


\begin{figure*}
\begin{center}
\includegraphics[width=0.8\textwidth]{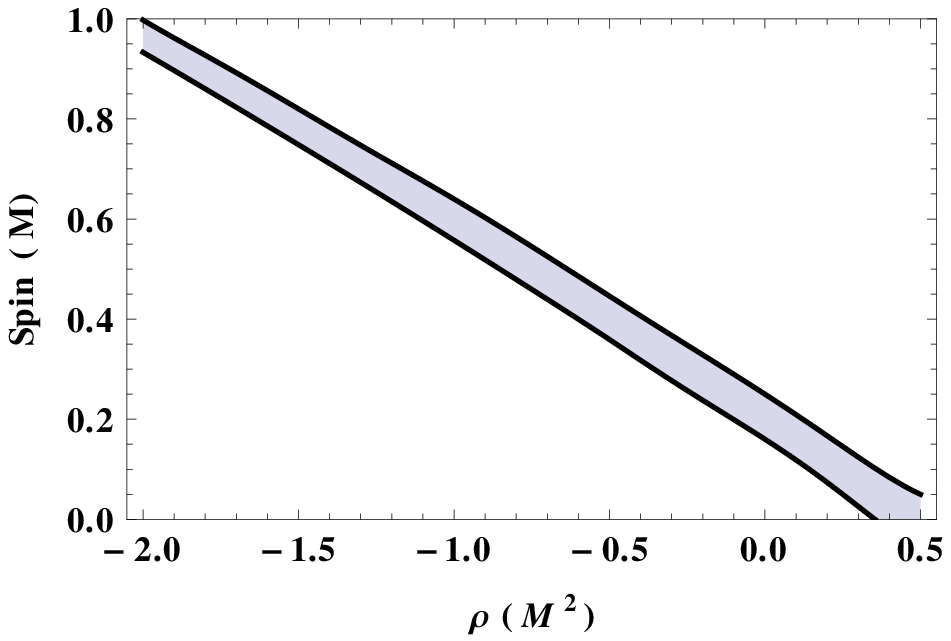}
\end{center}
\justifying
FIG. 4: Values of the spin $a$ and `charge' $\rho$ for the object GRO J$1655$-$40$ consistent with a measured mass of $M = 5.31 \pm 0.07 M_{\odot}$ and the observed QPO frequencies of $450$ Hz and $300$ Hz. The shaded region indicates the allowed zone.
\end{figure*}

\begin{figure*}
\begin{center}
\includegraphics[width=0.8\textwidth]{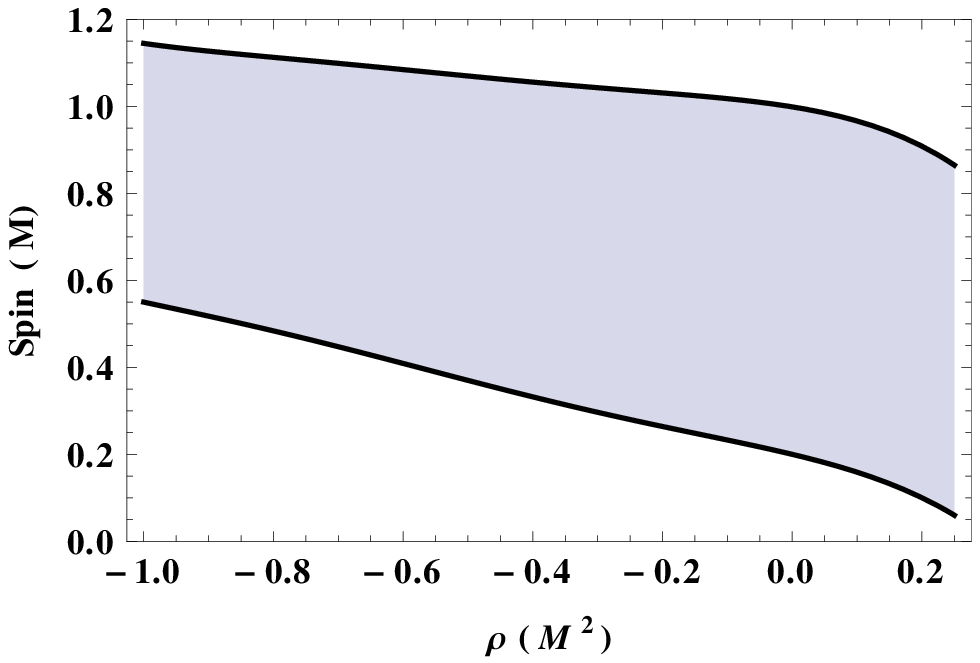}
\end{center}
\justifying
FIG. 5: Values of the spin $a$ and `charge' $\rho$ for the object GRS $1915$+$105$ consistent with a measured mass of $M = 14.0 \pm 4.4 M_{\odot}$. The large error bars in the mass mean that many possible values of $a$ and $\rho$ allow for the measured QPO frequencies of $67$ Hz and $35$ Hz. The shaded region indicates the allowed zone.
\end{figure*}

\section{Discussion}
In this paper we derive a set of mass and angular momentum multipole moments, that generalise earlier works to higher-order curvature theories of gravity, like $f(R)$ theories. The twist 4-vector $\boldsymbol{\omega}^{\dag}$ used by Hansen \cite{hansen} is generalized to ensure that there exists a scalar field $\omega$ such that $\omega_{\alpha} = \nabla_{\alpha} \omega$. The multipole moments then arise by evaluating moment functions, which satisfy Poisson equations. 

In $\S 5$ we calculate the multipole moments associated with a Kerr-Newman spacetime. Interestingly, the latter metric can arise either as a coupled Einstein-Maxwell solution of GR or as a pure vacuum solution to a range of $f(R)$ theories. Indeed, any GR solution with matter also arises as a vacuum $f(R)$ solution provided the trace of the stress energy tensor, $\Tr(\boldsymbol{T})$, is constant (as in the case of Maxwell fields in four spacetime dimensions) \cite{cruzd}. This has interesting consequences for the I-Love-Q type results regarding no-hair-like relations (or three-hair relations) among multipole moments for neutron stars and quark stars pioneered by Yunes, Yagi, and others \citep{yunyag,yagi1,stein}. Suppose we make a multipole measurement of a compact object, and it disagrees with the Kerr prediction. In GR, this would indicate the presence of an inherently hairy object such as a neutron star with strong magnetic field \cite{haskell}. However, it is also possible the object is actually a non-hairy black hole but that the theory of gravity is not GR. Higher-order moment measurements may remove the ambiguity but not if both objects satisfy no-hair-type relations (cf. \cite{mars,ross}).

Our results demonstrate that accretion disk models of QPOs, specifically the diskoseismology and kinematic resonance models, may be unable to uniquely identify properties of black holes or refute the no-hair theorem when combined with QPO data. We find that a wide range of possible combinations of the spin and `charge' are possible to explain QPO observations of GRS $1915$+$105$ and GRO J$1665$-$40$. Interestingly, the $g$-mode frequency for the Kerr case is bounded above due to $a \leq M$, while for the Kerr-Newman case we have a free parameter $\rho$ which, when large, allows for QPOs with kHz frequencies within the diskoseismology paradigm, such as those observed in low mass X-ray binaries like Sco X-1 \cite{scox1pap}.

Suppose we consider an axisymmetric spacetime that is not stationary. One can separate out the azimuth terms by constructing a spacelike Killing vector $\boldsymbol{\zeta} = \frac {\partial} {\partial \phi}$, allowing the construction of $S$ with respect to the spacelike Killing vector $\boldsymbol{\zeta}$ instead of $\boldsymbol{\xi}$. It would be worth exploring this avenue for generalizing the moment calculations in future work. For example, one could calculate the multipole moments associated with a radiating Vaidya spacetime, which appears as a vacuum solution to some $f(R)$ theories \cite{vaidya}. Additionally, one can apply the results to metrics other than Kerr-Newman, such as those derived using point-like Lagrangian techniques \cite{capnoe}. In GR, results such as those of Xanthopoulous and Gursel \cite{xan,gurs} demonstrate that each spacetime carries with it a unique set of multipole moments and vice versa. Furthermore, it can be shown that any stationary, asymptotically flat solution to Einstein's equations approaches the Kerr solution near infinity \cite{beigsim,simonbeig}. Since the Geroch-Hansen moments are evaluated at infinity, the two results together show that the Kerr metric is unique in GR. The extension of the moments presented here may help to prove uniqueness results in modified gravity along these lines.

\begin{acknowledgements}
We thank the anonymous referee for their carefully considered comments which significantly improved the quality of this manuscript. This work was supported in part by an Australian Postgraduate Award.
\end{acknowledgements}

\appendix

\section{Tensor relations on $S$}
In this section we derive some useful relations for tensor fields over $S$. First, it was shown by Geroch \cite{gerochgen} that there is a one-to-one correspondence between tensor fields $\mathscr{T}^{b\cdots d}_{a\cdots c}$ on $S$ and tensor fields $T^{b\cdots d}_{a\cdots c}$ on $M$. Hence the covariant derivative $D$ over $S$ can be written in terms of the covariant derivative $\nabla$ over $M$ as \cite{gerochgen,gerochgen2}
\begin{equation} \label{eq:covdv}
D_{e} T^{b \cdots d}_{a \cdots c} = h^{p}_{e} h^{m}_{a} \cdots h^{n}_{c} h^{b}_{r} \cdots h^{d}_{s} \nabla_{p} T^{r \cdots s}_{m \cdots n},
\end{equation} 
where $\boldsymbol{h}$ is the 3-metric on $S$ as in \eqref{eq:papapet}. Let $\boldsymbol{k}$ be an arbitrary vector field over $S$. Then using equation \eqref{eq:covdv} we find
\begin{equation}
\begin{aligned}
D_{a} D_{b} k_{c} =& h^{p}_{a}h^{s}_{b} h^{t}_{c} \nabla_{p} \nabla_{s} k_{t} - \lambda^{-1} h^{p}_{a} h^{q}_{b} h^{r}_{c} \left( \nabla_{p} \xi_{q} \right) \xi^{s} \nabla_{s} k_{r} \\
&-\lambda^{-1} h^{p}_{a} h^{q}_{b} h^{r}_{c} \left( \nabla_{p} \xi_{r} \right) \xi^{t} \nabla_{q} k_{t}.
\end{aligned}
\end{equation}
By antisymmetrizing the above equation to find $\left( D_{a} D_{b} - D_{b} D_{a} \right) k_{c}$, we obtain an expression for the Riemann tensor $\mathscr{R}^{a}_{bcd}$ over $S$:
\begin{equation} \label{eq:riem}
\hspace{-0.25cm}\mathscr{R}_{abcd} = h^{p}_{[a} h^{q}_{b]} h^{r}_{[c}h^{s}_{d]} \Big[ R_{pqrs} + 2  \lambda^{-1} (\nabla_{p} \xi_{q}) (\nabla_{r} \xi_{s}) + 2\lambda^{-1} (\nabla_{p} \xi_{r}) (\nabla_{q} \xi_{s}) \Big] .
\end{equation}
Contracting indices and simplifying the result using the definitions \eqref{eq:twist} and \eqref{eq:qrelation} we find that the Ricci tensor on $S$ $\mathscr{R}_{ab} = \mathscr{R}_{acb}^{c}$ satisfies

\begin{equation} \label{eq:ric}
\begin{aligned}
\hspace{-0.55cm}\mathscr{R}_{a b} =& \frac {1} {2} \lambda^{-2} \Big[ D_{a} \omega D_{b} \omega - h_{a b} D^{m} \omega D_{m} \omega \Big]\\
 &+ \frac {1} {2} \lambda^{-1}  D_{a} D_{b} \lambda - \frac {1} {4} \lambda^{-2} (D_{a} \lambda) (D_{b} \lambda) + h^{m}_{a} h^{n}_{b} R_{mn}  \\
&+ \frac {1} {2} \lambda^{-2} \left[ Q_{a} Q_{b} - D_{a} \omega Q_{b} - Q_{a} D_{b} \omega - h_{a b} Q^{m} Q_{m} + 2 h_{a b} Q^{m} D_{m} \omega \right] .
\end{aligned}
\end{equation}

From the definitions \eqref{eq:norm} and \eqref{eq:gentwist22} for $\lambda$ and $\omega$, we find that these objects satisfy the Laplace equations [using \eqref{eq:covdv}]

\begin{equation} \label{eq:omega}
D^{a} D_{a} \omega = \frac {3} {2} \lambda^{-1} D_{m} \omega D^{m} \lambda - \frac {3} {2} \lambda^{-1} Q_{m} D^{m} \lambda + D^{a} Q_{a}  ,
\end{equation}
and
\begin{equation} \label{eq:lambda}
\begin{aligned}
\hspace{-0.95cm}D^{j} D_{j} \lambda =& \frac {1} {2} \lambda^{-1} (D^{m} \lambda) (D_{m} \lambda)  \\
\hspace{-0.95cm}&-  \lambda^{-1} \big[ D^{m} \omega D_{m} \omega - 2 D^{m} \omega Q_{m} + Q^{m} Q_{m} \big] - 2 R_{m n} \xi^{m} \xi^{n}  .
\end{aligned}
\end{equation}
The fundamental field equations on $S$ are given by \eqref{eq:omega} and \eqref{eq:lambda} in combination with a set of field equations for $h_{ij}$ will depend on the specific theory of gravity. As an example, the vacuum field equations for $f(R)$ gravity over $M$ read (e.g. \cite{felice})
\begin{equation} 
f'(R) R_{\mu \nu} - \frac {1} {2} f(R) g_{\mu \nu} + (g_{\mu \nu} \square - \nabla_{\mu} \nabla_{\nu} ) f'(R) = 0  .
\end{equation}
Using the identities \eqref{eq:covdv} and \eqref{eq:ric}, the $f(R)$ field equations over $S$ read
\begin{widetext}
\begin{equation} \label{eq:fofrfieldeqns}
\hspace{-0.75cm}\begin{aligned}
&\Bigg\{ \frac {1} {2} \lambda^{-2} \Big[ D_{a} \omega D_{b} \omega - h_{a b} D^{m} \omega D_{m} \omega \Big] + \frac {1} {2} \lambda^{-1}  D_{a} D_{b} \lambda - \frac {1} {4} \lambda^{-2} (D_{a} \lambda) (D_{b} \lambda) + h^{m}_{a} h^{n}_{b} R_{mn} + \frac {1} {2} \lambda^{-2} \Big[ Q_{a} Q_{b} - D_{a} \omega Q_{b} - Q_{a} D_{b} \omega - h_{a b} Q^{m} Q_{m} \\
&+ 2 h_{a b} Q^{m} D_{m} \omega \Big] + h_{ab} D^{j} D_{j} - D_{a} D_{b} \Bigg\}  f'\left[ - \lambda^{-2} \left( \tfrac {3} {2} D^{m} \omega D_{m} \omega + \tfrac {3} {2} Q_{k} Q^{k} - 3 D_{p} \omega Q^{p} \right) + R_{mn} \left( h^{mn} - \tfrac {1} {\lambda} \xi^{m} \xi^{n} \right) \right] - \frac {1} {2} h_{a b} \\
& f\left[ - \lambda^{-2} \left( \tfrac {3} {2} D^{m} \omega D_{m} \omega + \tfrac {3} {2} Q_{k} Q^{k} - 3 D_{p} \omega Q^{p} \right) + R_{mn} \left( h^{mn} - \tfrac {1} {\lambda} \xi^{m} \xi^{n} \right) \right] = f'(\mathscr{R}) \mathscr{R}_{ab} - \frac {f(\mathscr{R})} {2} h_{ab} + \left( h_{ab} D^{j} D_{j} - D_{a} D_{b} \right) f'(\mathscr{R}) .\\
\end{aligned}
\end{equation}
\end{widetext}
The fundamental equations for an $f(R)$ gravity over $S$ with Killing vector $\boldsymbol{\xi}$ are \eqref{eq:omega}, \eqref{eq:lambda}, and \eqref{eq:fofrfieldeqns}. They reduce to those found by Geroch \cite{gerochcurvedmult} for $R_{\mu \nu} = 0$ and $f(R) = R$.

\section{Conformally invariant Poisson equations}

The wave equation for a scalar field $\phi$ with forcing term $f$ on a curved background,
\begin{equation} \label{eq:notconformal}
g^{\alpha \beta} \nabla_{\alpha} \nabla_{\beta} \phi = f,
\end{equation}
is not conformally invariant; a metric $\boldsymbol{g}$ scaled by some function $\psi$ (i.e. $\boldsymbol{\tilde{g}} = \psi^2 \boldsymbol{g}$) will not be a solution to \eqref{eq:notconformal} unless $\psi$ is trivial. In general, a conformally invariant equation will not exist unless $f$ is of conformal weight $-(1+n/2)$, i.e. $\boldsymbol{\tilde{g}} \rightarrow \psi^2 \boldsymbol{g} \implies f \rightarrow \psi^{-1-n/2} f$ \cite{koba}. To obtain a conformally invariant equation one must insert a term involving the Ricci scalar of the form (see \cite{wald}, page 447)
\begin{equation} \label{eq:conformal}
\left[ g^{\alpha \beta} \nabla_{\alpha} \nabla_{\beta} - \frac {n -2} {4 (n-1)} R \right] \phi = f ,
\end{equation}
where we have $n = \dim(M)$. On the space $S$ we have $n=3$ and as such equation \eqref{eq:conformal} reduces to
\begin{equation} \label{eq:somepoisss}
\left(h^{a b} D_{a} D_{b} - \frac {\mathscr{R}} {8}\right) \phi = f ,
\end{equation}
where $\boldsymbol{h}$ is the 3-metric as in \eqref{eq:papapet}. Equation \eqref{eq:somepoisss} is precisely the form of the Poisson equations \eqref{eq:poisson} considered in this paper, where $f = \tfrac{15}{16} \lambda^{-2} \kappa \phi$. This is the same function $f$ used by Hansen, where it was proved that $f$ is of conformal weight $-5/2$ \cite{hansen}, which is equal to $-1 - n/2$ for $n=3$. Therefore, equation \eqref{eq:poisson} is conformally invariant.

\section{The differential equation for $Q$}

In order to generalise the twist scalar $\omega$, we require the introduction of a one-form $\boldsymbol{Q}$ obeying
\begin{equation} \label{eq:qrelation2}
d \boldsymbol{Q} = - \star \left[ \boldsymbol{\xi} \wedge \boldsymbol{R} \left(\boldsymbol{\xi}\right)\right] ,
\end{equation}
where the differential equation is solved with Dirichlet boundary conditions to ensure $\boldsymbol{Q} = 0$ for $R^{\mu}_{\nu} \xi^{\nu} = 0$.  In a Boyer-Lindquist coordinate system $\{t,r,\theta,\phi\}$ we may take $\xi^{\mu} = \delta^{\mu}_{t}$ and $\boldsymbol{Q} = \left(0,Q_{r},Q_{\theta},Q_{\phi}\right)$ without loss of generality. Then \eqref{eq:qrelation2} can be written as the coupled system
\begin{equation} \label{eq:qrelation4}
\frac {\partial Q_{\phi}} {\partial \theta} - \frac {\partial Q_{\theta}} {\partial \phi} = 2 \sqrt{-g} R^{r}_{t} , 
\end{equation}
\begin{equation} \label{eq:qrelation5}
\frac {\partial Q_{r}} {\partial \phi} - \frac {\partial Q_{\phi}} {\partial r} =  2 \sqrt{-g} R^{\theta}_{t} ,
\end{equation}
\begin{equation} \label{eq:qrelation6}
\frac {\partial Q_{r}} {\partial \theta} - \frac {\partial Q_{\theta}} {\partial r} = -2 \sqrt{-g} R^{\phi}_{t} ,
\end{equation}
where we make use of the symmetry $\Gamma^{\alpha}_{\mu \nu} = \Gamma^{\alpha}_{\nu \mu}$ of the Christoffel symbols. In practice, the system \eqref{eq:qrelation4}--\eqref{eq:qrelation6} can be solved through repeated integration (e.g. \cite{john}).

\end{document}